\documentclass[aps,prc,preprint]{revtex4}
\usepackage{graphicx}
\def\be{\begin{equation}}
\def\ee{\end{equation}}
\def\bea{\begin{eqnarray}}
\def\eea{\end{eqnarray}}
\def\beq{\begin{equation}}
\def\eeq{\end{equation}}
\def\nn{\nonumber}

\def\mv{{\mathbf v}}
\def\mK{{\mathbf K}}
\def\mr{{\mathbf r}}
\def\mq{{\mathbf q}}
\def\mk{{\mathbf k}}
\begin{document}
\title{Back-to-back Correlations for Finite Expanding Fireballs}
\author{Sandra S. Padula, G. Krein}
\affiliation{Instituto de F\'{\i}sica Te\'orica,
Universidade Estadual Paulista\\
Rua Pamplona 145, 01405-900 S\~ao Paulo, SP - Brazil}
\author{T. Cs\"org\H o}
\affiliation{MTA KFKI RMKI, H - 1525 Budapest 114, POBox 49,
Hungary}
\author{Y. Hama}
\affiliation{Instituto de F\'{\i}sica, Universidade de S\~ao
Paulo, Caixa Postal 66318, 05389-970 S\~{a}o Paulo, SP - Brazil}
\author{P. K. Panda}
\affiliation{Departamento de F\'\i sica-CFM, Universidade Federal
de Santa Catarina \\ Caixa Postal 476, 88040-900 Florian\'opolis,
SC - Brazil}
\begin{abstract}
Back-to-Back Correlations of particle-antiparticle pairs are
related to the in-medium mass-modification and squeezing of the
quanta involved. They are predicted to appear when hot and dense
hadronic matter is formed in high energy nucleus-nucleus
collisions. The survival and magnitude of the Back-to-Back
Correlations of boson-antiboson pairs generated by in-medium mass
modifications are studied here in the case of a thermalized,
finite-sized, spherically symmetric expanding medium. We show that
the BBC signal indeed survives the finite-time emission, as well
as the expansion and flow effects, with sufficient intensity to be
observed at RHIC.
\end{abstract}

\date{\today}

\maketitle

\section{Introduction}
Recently, it has been shown~\cite{ac,acg} that large back-to-back
correlations (BBC) of particle-antiparticle pairs of bosonic
particles might appear in high energy nucleus-nucleus collisions
as a consequence of in-medium mass modification of the bosons.
Detailed calculations indicate that the BBC signal appears for
values of transverse momenta below 1-2 GeV/c. More recently, it
was shown~\cite{fbbc} that BBC of similar strength might appear
for fermionic particles as well. The main physical ingredient used
in the evaluation of the effects of in-medium modified masses on
two particle correlation functions is a quantum-mechanical
correlation induced by a nonzero overlap between in-medium and
free states. The induced quantum mechanical correlation can be
represented in terms of two-mode squeezed states of the
asymptotic, observable states and is implemented through a
Bogoliubov-Valatin transformation.

The possibility of measuring a significant BBC signal in heavy-ion
collisions opens new interesting possibilities for accessing the
properties of the matter formed in such collisions. The BBC signal
is linked to in-medium mass modifications of hadrons in the hot
and dense environment the detected particles experience before
freezing out and in this sense BBC measurements provide
independent pieces of information on medium modifications from the
ones obtained from dilepton yields and spectra. However, there are
several additional physical effects that interfere with mass
modifications of the detected particles in the interpretation of
the BBC signal.

All studies in Refs.~\cite{ac}, \cite{acg} and \cite{fbbc} were
restricted to infinite, static media. For testing the robustness
of the effect, we generalize the previous studies to a more
realistic situation of mass modification in a finite-sized,
expanding thermalized medium. In this first investigation of such
effects, we use a simple hydrodynamical model for the expansion
and simple three-dimensional Gaussian profile for the size of the
system. Although simple, the model is rich enough to indicate the
influence of the expansion and finite-size effect on BBC.

In the next section we review the basic ingredients of the model
and in Section~\ref{sec:3} we generalize the model to include
expansion effects. Numerical results are presented in
Section~\ref{sec:numeric} and Section~\ref{sec:concl} presents our
Conclusions and Future Perspectives.

\section{Review of The Model - Infinite Homogeneous Medium}
\label{sec:2}

In the present paper we concentrate on bosonic BBC and restrict
the discussion to cases where the boson is its own antiparticle --
like the $\phi$ meson. We are interested in the two-particle
correlation function
\be C_2({\mk}_1,{\mk}_2) = \frac{N_2(\mk_1,\mk_2)} {N_1(\mk_1) \,
N_1(\mk_2) }, \label{defC2} \ee
where $N_1(\mk_i)$ and $N_2(\mk_1,\mk_2)$ are, respectively, the
invariant single-particle and two-particle momentum distributions
\begin{eqnarray}
N_1(\mk_1) &=& \omega_{\mk_1} \frac{d^3N}{d\mk_1} =
\omega_{\mk_1}\,
\langle a^\dagger_{\mk_1} a_{\mk_1} \rangle , \\
N_2(\mk_1,\mk_2) &=& \omega_{\mk_1} \omega_{\mk_2} \, \langle
a^\dagger_{\mk_1} a^\dagger_{\mk_2} a_{\mk_2} a_{\mk_1} \rangle
\nonumber \\
&=& \omega_{\mk_1} \omega_{\mk_2} \, \Bigl[\langle
a^\dagger_{\mk_1} a_{\mk_1}\rangle \langle a^\dagger_{\mk_2}
a_{\mk_2} \rangle + \langle a^\dagger_{\mk_1} a_{\mk_2}\rangle
\langle a^\dagger_{\mk_2} a_{\mk_1} \rangle + \langle
a^\dagger_{\mk_1} a^\dagger_{\mk_2} \rangle \langle a_{\mk_2}
a_{\mk_1} \rangle\Bigr], \label{rand}
\end{eqnarray}
where $a^\dagger_\mk$ and $a_\mk$ are free-particle creation and
annihilation operators of scalar quanta, and the angular brackets
mean thermal averages. The factorization of the expectation value
of four operators into products of expectation values of two
operators in Eq.(\ref{rand}) has been derived as a generalization
of Wick's theorem for {\em locally} equilibrated (chaotic) systems
in Refs.\cite{sm,gykw,surp}. Introducing the chaotic and squeezed
amplitudes as
\bea G_c({\mk_1},{\mk_2}) &=& \sqrt{\omega_{\mk_1} \omega_{\mk_2}} \; \langle
a^\dagger_{\mk_1} a_{\mk_2} \rangle,
\label{gc}\\
G_s({\mk_1},{\mk_2}) &=& \sqrt{\omega_{\mk_1} \omega_{\mk_2} } \; \langle
a_{\mk_1} a_{\mk_2} \rangle, \label{gs} \eea
the two-particle correlation function can be written as
\be
C_2({\mk_1},{\mk_2}) = 1 + \frac{|G_c({\mk_1},{\mk_2})|^2}{G_c({\mk_1},{\mk_1})
G_c({\mk_2},{\mk_2})} + \frac{|G_s({\mk_1},{\mk_2}) |^2}{G_c({\mk_1},{\mk_1})
G_c({\mk_2},{\mk_2}) }.
\label{fullcorr}
\ee
The $G_c(1,2)$ is the usual Hanbury-Brown-Twiss (HBT) amplitude
and $G_s(1,2)$ is BBC amplitude.

The thermal average of an operator $\hat{O}$, $\langle \hat{O}
\rangle = {\rm Tr} ( \hat{\rho}\,\hat{O})$, is calculated with a
density matrix $\hat{\rho}$ corresponding to medium-modified,
thermalized quanta. The crucial point is that the in-medium
thermalized quanta {\em are not} the ones detected. The detected
quanta have energy-momentum $k^{\mu} = (\omega_\mk, \mk)$,
$\omega_{\mk}^2 = { \mathbf k}^2 + m^2$ and are described by the
creation and annihilation operators $a^\dagger_\mk$ and $a_\mk$.
However, if we denote by $b^\dagger_\mk$ and $b_\mk$ the creation
and annihilation operators of in-medium, thermalized quanta with
$k^{* \mu} = (\Omega_\mk, \mk)$, $\Omega^2_\mk= \mk^2 +
m^2_*(|\mk|)$, we can relate the $(a^\dagger_{\mathbf k}, a_\mk)$
to $(b^\dagger_\mk, b_\mk)$ through a Bogoliubov-Valatin (BV)
transformation. Specifically, the annihilation operator
$a_{\mk_1}$ for the asymptotic quanta with momentum $\mk_1$ is
related to the in-medium operators $b_{\mk_1}$ and
$b^\dagger_{\mk_1}$  as~\cite{ac}:
\be
a_{\mk_1} = c_{\mk_1} \, b_{\mk_1} + s^*_{-\mk_1} \,
b^\dagger_{-\mk_1} \equiv C_1 + S_{-1}^\dagger, \label{asq} \ee
where we have introduced the notation $C_1 = c_{\mk_1} \,
b_{\mk_1}$ and  $S_{-1} = s^*_{-\mk_1} \, b^\dagger_{-\mk_1}$ to
simplify later notation, and
\begin{equation}
c_{\mk} = \cosh[f_{\mk}], \hspace{0.5cm}s_{\mk} = \sinh[f_{\mk}],
\hspace{0.5cm} f_{\mk} = \frac{1}{2} \log
\left(\frac{\omega_{\mk}}{\Omega_{\mk}}\right) .
\label{sqr}
\end{equation}
The BV transformation for the creation operator
$a^\dagger_{\mk_1}$ is obtained from Eq.~(\ref{asq}) by Hermitean
conjugation. As is well known, the Bogoliubov transformation is
equivalent to a squeezing operation, and this motivates calling
$f_\mk$ the mode-dependent squeezing parameter. In this way, it is
the squeezing parameter $f_\mk$ that carries the in-medium
effects. Using the BV relation, we obtain for the thermal averages
in Eqs.~(\ref{gc}) and (\ref{gs})
\begin{eqnarray}
G_c({\mk_1},{\mk_2}) &=& \sqrt{\omega_{\mk_1} \omega_{\mk_2}} \; \left[\langle
C^\dagger_1 C_2\rangle + \langle S_{-1}
S^\dagger_{-2}\rangle\right],
\label{gc1}\\
G_s({\mk_1},{\mk_2})&=& \sqrt{\omega_{\mk_1} \omega_{\mk_2}} \; \left[ \langle
S^\dagger_{-1} C_2 \rangle + \langle C_1 S^\dagger_{-2} \rangle
\right].
\label{gs1}
\end{eqnarray}

After performing the thermal averages indicated above, with the
help of a thermal density matrix $\hat{\rho}$ corresponding to the
in-medium modified, thermalized quanta, the resulting expressions
for the case of an homogeneous medium are
\begin{eqnarray}
    G_c(1,2)  &=& \left\{ \frac{E_{1,2}}{(2 \pi)^3 }
    \left[ |c_{1,2}|^2
\, n_{1,2} + |s_{-1,-2}|^2 \,  (n_{-1,-2} + 1) \right] \right\}
       ~ V \delta_{1,2}, \label{e:gchom}\\
    G_s(1,2) & = & \Bigl\{\frac{E_{1,2}}{(2 \pi)^3 }\left[
   s^*_{-1,2}\, c_{2,-1} \, n_{-1,2} + c_{1,-2} \, s^*_{-2,1}
   \left(n_{1,-2} + 1\right) \right] \Bigr\}~ V \delta_{1,-2} , \label{e:gshom}
\end{eqnarray}

From Eq. (\ref{e:gchom}) and (\ref{e:gshom}) it is easily seen
that, in the approximation of a sudden freeze out, and in the case
of a homogeneous medium, $G_c({\mk_1},{\mk_2})\propto V
\delta_{1,2}$ and $G_s({\mk_1},{\mk_2}) \propto V \delta_{1,-2}$.
Therefore, the amplitudes $G_c({\mk_1},{\mk_2})$ and
$G_s({\mk_1},{\mk_2})$ are non-vanishing only for $\mk_1 = \mk_2$
and $\mk_1 = - \mk_2$ respectively. In the expression for the
two-particle correlation function the volume factors cancel out,
and we obtain~\cite{acg}
\bea
C_2(\mk, \mk)  &=& 2, \\
C_2(\mk, -\mk)  &=&  1 + \frac{|c_{\mk} s^*_{\mk} n_{\mk} +
c_{-\mk} s^*_{-\mk} \left(n_{-\mk} + 1\right)|^{\, 2} } {n_1(\mk)
\, n_1(-\mk)}, \label{e:c2b} \eea
where $n_1(\mk)$ is defined by
\be
N_1(\mk) = \frac{V}{(2 \pi)^3} \, \omega_\mk \, n_1(\mk), \\
\ee
with
\begin{equation}
n_1(\mk) = \left[|c_{\mk}|^2 n_{\mk} + |s_{-\mk}|^2 \left(n_{-\mk}
+ 1 \right)\right],
\end{equation}
and $n_\mk$ is the Bose-Einstein distribution function of the
in-medium quanta with energy $\Omega_\mk$ at temperature $T$. The
exact value of the intercept, $C_2(\mk,\mk)=2$, is a
characteristic signature  of a chaotic Bose gas without dynamical
2-body correlations outside the domain of Bose-Einstein
condensation.

 We should note that Eq.~(\ref{e:c2b}) is valid {\em
only} in the rest frame of the medium, i.e., {\it the correlation
is back-to-back only in the rest frame of the matter}. In the next
section we extend the model to a medium with finite size
corresponding to a fireball, which is exploding with a position
dependent flow velocity field distribution, so that only the
central point of this exploding fireball is at rest in the frame
of the observation.

\section{Spectra and correlations for mass-shifted bosons in
finite expanding  systems} \label{sec:3}

We are mainly interested here in the study of the squeezed
correlation function --  first and third terms of
Eq.~(\ref{fullcorr}). For studying the expansion of the system we
adopt for the emission function the non-relativistic
hydrodynamical parameterization of Ref.~\cite{Csorgo:fg}, which
was shown later to actually be a non-relativistic hydrodynamical
solution. In this model the fireball expands in a spherically
symmetric manner with non-relativistic four-velocity $u^\mu =
\gamma \, (1, {\mathbf v})$, with $\gamma = (1-{\mathbf
v}^2)^{-1/2} \approx 1 +{\mathbf v}^2/2$, where
$${\mathbf v} =
\langle u\rangle {\mathbf r}/R,$$ $\langle u\rangle$ and $R$ are,
respectively, the mean expansion velocity and the radius of the
fireball. Thus, we divide the inhomogeneous medium into
independent cells and assume that Eqs.~(\ref{gc1}) and (\ref{gs1})
can be evaluated locally within each cell using the BV
transformation of Eq.~(\ref{asq}) -- and its Hermitian conjugate.
Then, the amplitudes $G_c$ and $G_s$ can be written in the special
form derived by Makhlin and Sinyukov~\cite{sm}, which are given by
Eqs.~(22) and (23) of Ref.~\cite{acg}, namely
\begin{eqnarray}
G_c({\mk_1},{\mk_2}) &=& \int \frac{d^4\sigma_{\mu}(x)}{(2 \pi)^3}
\, K^\mu_{1,2} \, e^{i \, q_{1,2} \cdot x} \, \Bigl\{|c_{1,2}|^2
\, n_{1,2}(x) + |s_{-1,-2}|^2 \,  \left[n_{-1,-2}(x) + 1\right]
\Bigr\}, \label{e:gcinhom} \\
G_s({\mk_1},{\mk_2})  &=& \int \frac{d^4\sigma_{\mu}(x)}{(2
\pi)^3} \, K^\mu_{1,2} \, e^{2 \,i \, K_{1,2} \cdot x} \Bigl\{
s^*_{-1,2}\, c_{2,-1} \, n_{-1,2}(x) + c_{1,-2} \, s^*_{-2,1} \,
\left[n_{1,-2}(x) + 1\right] \Bigr\}. \nonumber\\
\label{e:gsinhom}
\end{eqnarray}
Here $d^4\sigma^{\mu}(x) = d^3\Sigma^{\mu}(x;\tau_f)\, F(\tau_f)
d\tau_f$ is the product of the normal-oriented volume element
depending parametrically on the freeze-out hypersurface parameter
$\tau_f$ and  on its invariant distribution function $F(\tau_f)$.
We should notice that, in the particular case in which each
$d^4\sigma_{\mu}(x)$ of Eqs. (17) and (18) is parallel to $u^\mu$,
that is, the emission from an elementary cell mentioned above
occurs instantaneously in its proper frame, the exponential factor
there will give rise, upon integration over the cell assuming it
is large enough, to the same factor $\delta_{1,2}$ or
$\delta_{1,-2}$ that were  present in Eqs. (11) or (12),
respectively. As mentioned at the end of Sec. II, the arguments of
$\delta$ here are not {\bf k}$_1$ and {\bf k}$_2$ of the left-hand
side, but should be understood as given in the proper frame of the
cell. In what follows, the condition of instantaneous emission in
the proper frame of each cell is assumed to be approximately
verified, since our calculation is non-relativistic. However, we
should remark that due to the fact that our elementary cells are
not always large, the correlation described above is only
approximately back-to-back.

\begin{figure}[htb]
%\resizebox{25pc}{!}{\includegraphics{figs_vetsFab_crop.eps}}
\resizebox{25pc}{!}{\includegraphics{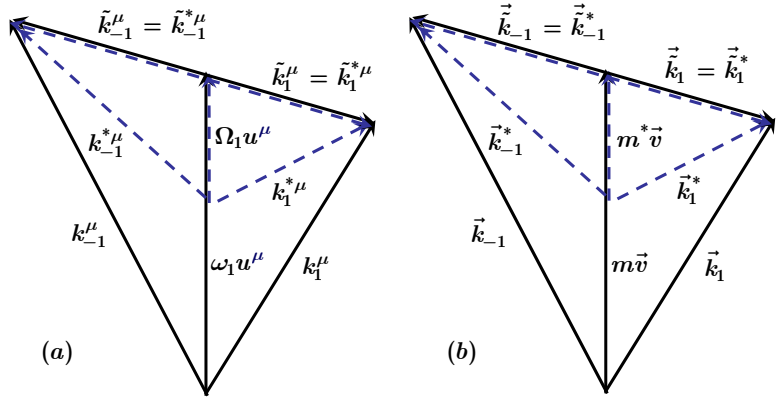}}
\caption{The definition of the various momentum possibilities
discussed above (where the index 1 was chosen for simplicity), and
the illustration of the dependence of the back-to-back momentum
pairs on the four-velocity and on the mass-shift, is shown in (a)
using relativistic notation. Solid lines represent the locally
back-to-back momentum pair  and its components for asymptotic
quanta; dashed lines represent the same for in-medium modified,
mass-shifted quanta. In (b), the analogous situation is
illustrated using the non-relativistic notation. }
\label{vectors}\vskip0.5cm
\end{figure}

The other quantities appearing in Eq. (\ref{e:gcinhom}) and
(\ref{e:gsinhom}) are
$n_{i,j}(x) \equiv n(x,K_{i,j})$, the local density distribution,
and $c_{i,j} = \cosh[f_{i,j}(x)]$ and $s_{i,j} =
\sinh[f_{i,j}(x)]$, squeezed functions with
\begin{equation}
f_{i,j}(x)  = \frac{1}{2}\log \left[\frac{ K^\mu_{i,j}(x) \, u_\mu
(x)} {K^{*\,\nu}_{i,j}(x) \, u_\nu(x) } \right],
\end{equation}
where $u^\mu(x)$ is the local flow vector at freeze-out. The
relative and the average pair four-momentum coordinates are
defined as $q^0_{1,2}= \omega_1 - \omega_2$, $\mq_{1,2} = \mk_1 -
\mk_2$,  $K^0_{1,2} = (\omega_1 + \omega_2)/2$, and $\mK_{1,2} =
(\mk_1 + \mk_2)/2$. Also, we identify in-medium and squeezed
quantities by superscripted asterisks. The relative
$q^\mu_{i,j}(x)$ and total four momenta $K^\mu$ of particles 1 and
2 are given by
\begin{equation}
q^{\mu}_{i,j}(x) = k^{\mu}_i(x) - k^{\mu}_j(x),
\hspace{1.0cm}K^{\mu}_{i,j}(x) = \frac{1}{2} \left[k^{\mu}_i(x) +
k^{\mu}_j(x) \right] ,
\end{equation}
where $k^\mu_i(x)$ for $i= \pm 1,\pm 2$ are given by
\begin{equation}
k^\mu_{\pm i}(x) = \omega_{\mk_i}(x) \, u^\mu (x) \pm \tilde
k^\mu_i(x), \hspace{1.0cm} \omega_{\mk_i}(x) = \sqrt{m^2 -
\tilde{k_i}^{\mu}\tilde{k_i}_\mu }  = k^\mu_i \, u_\mu(x) ,
\end{equation}
with $\tilde k^{\mu}_i$ orthogonal to~$u^\mu(x)$:
\be
\tilde k^{\mu}_i = k^{\mu}_i -  k_i \cdot u(x)\, u^\mu(x),
\label{tildek}
\ee
The corresponding in-medium quantities are given by
\begin{equation}
q^{*\mu}_{i,j}(x) = k^{*\mu}_i(x) - k^{*\mu}_j(x),
\hspace{1.0cm}K^{*\mu}_{i,j}(x) = \frac{1}{2} \left[k^{*\mu}_i(x)
+ k^{*\mu}_j(x) \right] ,
\end{equation}
and
\begin{equation}
k^{*\mu}_{\pm i}(x) = \Omega_{\mk_i}(x) \; u^\mu(x) \pm \tilde
k^{*\mu}_i(x),\hspace{1.0cm}\Omega_{\mk_i}(x) = \sqrt{m^2_*(x,
\tilde{k}) - \tilde{k}^{*\mu}_i \tilde{k}^{*}_{i\mu} } =
k^{*\mu}_i u_\mu(x),
\end{equation}
with
\be \tilde k^{*\mu}_i = k^{*\mu}_i -  k^*_i \cdot u(x)\, u^\mu(x).
\label{tildek*}
\ee
Now, it is not difficult to show that $\tilde{k}^{*\mu}_i (x) =
\tilde{k}^\mu_i (x)$ and therefore no star is necessary in
$\tilde{k}$ for the in-medium quantities. It should be noted that
this equality was not clearly emphasized in Ref. \cite{acg}. An
important aspect of these relations is that due to the mass
modification, the energy in the local co-moving frame is modified
from $\omega_{\mk_i}(x) = k^\mu_i(x) \, u_\mu(x) = k^\mu_{\pm
i}(x)\, u_\mu(x)$ to $\Omega_{\mk_i}(x) = k^{*\mu}_i(x) \,
u_\mu(x) = k^{*\mu}_{\pm i}(x) \, u_\mu(x)$, without modifying the
component of the four-momentum orthogonal to the four-velocity.
The above definitions are the detailed write-up of similar
definitions of Ref.~\cite{acg}, where a more succinct notation has
been used and the misprint signs ($\mp$, instead of $\pm$) in Eq.
(27) and (28) of Ref.\cite{acg} have been corrected, respectively,
in expressions for $k^\mu_{\pm i}(x)$ and $k^{*\mu}_{\pm i}(x)$
above.
These definitions of momenta are illustrated in
Fig.~\ref{vectors}, corresponding to the relativistic and
non-relativistic limits, in parts (a) and (b), respectively.

Using the above expressions, the squeezing parameter can be
evaluated as
\begin{equation}
f_{i,j}(x)=\frac{1}{2}\log\left[\frac{K^{\mu}_{i,j}(x)\, u_\mu
(x)} {K^{*\nu}_{i,j}(x) \, u_\nu(x)}\right] = \frac{1}{2}
\log\left[\frac{\omega_{\mk_i}(x) + \omega_{\mk_j}(x)}
{\Omega_{\mk_i}(x) + \Omega_{\mk_j}(x)}\right] \equiv f_{\pm i,\pm
j}(x), \label{e:rxk}
\end{equation}
where the spatial coordinate dependence enters through the position
dependence of either the four-velocity or the in-medium mass modification,
or both. However, it does not matter which of the locally back-to-back
momenta are used for the evaluation of the amount of squeezing.

In this paper, we focus on the effects of expansion on the
back-to-back correlations - or, with other words, does the flow
wash out the signal for these correlations or not? Although the
formalism of the squeezed back-to-back correlations was worked out
with expanding systems in Ref. 2, no detailed investigations were
performed to quantitatively study e.g. the strength of the signal
with varying the strength of the flow.

Here we intend to investigate this question in one of the simplest
geometrical cases. For the sake of clarity, we evaluate the flow
effects for a non-relativistically expanding, spherically
symmetric fireball, that freezes out at a constant temperature $T$
and has a Gaussian density profile. In this sense, we adopt the
model emission function of Ref.\cite{Csorgo:fg}, that was
developed to study single particle spectra and Bose-Einstein (HBT)
correlation functions in the simplest possible case of expanding
systems. Later on it has been realized, that this emission
function corresponds to the simplest member of a new family of
exact solutions of non-relativistic hydrodynamics\cite{Csorgo:c},
which can be generalized in a straightforward manner to
cylindrically and ellipsoidally symmetric\cite{Csorgo:d}
expansions, as well as to the case of relativistic expansions
\cite{Csorgo:e}, and in all cases, systems that expand with
inhomogeneous temperature profiles \cite{Csorgo:d,Csorgo:f}.

Before investigating in detail the effects of various kind of
inhomogeneities in the flow profiles and in the temperature
profiles, let us turn our attention to the non-relativistic limit
adopting the simplest possible scenario for the expansion that
leads to analytic forms \cite{Csorgo:fg,Csorgo:c}.

For the sake of clarity, we present the explicit expressions in
the non-relativistic limit of the above quantities, which is the
appropriate limit for our non-relativistic flow model. Writing
$u^\mu = \gamma (1,{\mathbf v})$ and using $\gamma = (1-{\mathbf
v}^2)^{-1/2} \approx 1+{\mathbf v}^2/2 \approx 1$, we have that
Eq.~(\ref{tildek}) leads to
\be \mk_i - m \mv = \mk^*_i - m_* \mv, \ee
and therefore
\bea \mk_{\pm i}(\mr) &=& m\,\mv(\mr) \mp m\, \mv(\mr) \pm \mk_i ,
\\
\mk^*_{\pm i}(\mr) &=& m_*\,\mv(\mr) \mp m_*\, \mv(\mr) \pm
\mk^*_i. \eea
With this, we obtain
\be K^{*\mu}_{i,j}(x) \, u_\mu \approx  m_{*} + \frac{1}{2m_{*}}
\left\{ [{\mathbf K}^{*}_{i,j}- m_{*}{\mathbf v}({\mathbf r})]^2 +
\frac{1}{4} (\mq^*_{i,j})^2 \right\},
\label{Kstar}
\ee
where the total and relative local in-medium momenta are given by
\smallskip
\begin{eqnarray}
\mK^*_{1,2}(\mr) = \frac{1}{2}(\mk_1+ \mk_2) + (m_* - m) \, \mv
(\mr)\,,
&& \mq_{1,2}^{*}(\mr) = (\mk_1- \mk_2) \\
\mK^*_{1,-2}({\mathbf r})  = m_* \mv(\mr) + \frac{1}{2} (\mk_1-
\mk_2) \,,
&& \mq^*_{1,-2}(\mr) = -2 m {\mathbf v}({\mathbf r}) +
(\mk_1 + \mk_2) \\
\mK^*_{-1,2}(\mr) = m_* \mv(\mr) - \frac{1}{2}(\mk_1 - \mk_2) \,,
&& \mq^*_{-1,2}(\mr) = 2 m \mv(\mr) - (\mk_1 + \mk_2)\\
\mK^*_{-1,-2}(\mr)  = (m_* + m) \mv(\mr) -\frac{1}{2} ( \mk_1 +
\mk_2 ) \,, && \mq^*_{-1,-2}(\mr) = -(\mk_1 - \mk_2).
\end{eqnarray}
\medskip
The unstarred $\mK_{i,j}$ and $\mq_{i,j}$ momenta are obtained
from the above by replacing $m_*$ by $m$ in these expressions.
Note that $\mq^*_{i,j} = \mq_{i,j}$, as it should be. These
relations imply, that in the non-relativistic limit,
\be \mK^*_{ij} - m_* \mv = \mK_{ij} - m \mv.
\label{nonrelK}
\ee

In discussing finite-size effects, we distinguish between the
volume of the entire thermalized medium, denoted by $V$, and the
volume filled with mass-shifted quanta, denoted by $V_s$.
Naturally, $V_s \leq V$ in the general case. In the derivation of
the expressions for $G_c(1,2)$ and $G_s(1,2)$, for simplicity, we
introduce a Gaussian profile function in the integrands, i.e., we
consider that the volumetric region where the mass $m_*$ is
significantly modified is smooth and Gaussian in shape. In other
words, instead of considering a particular domain of integration,
we perform the spatial integrals for $G_c(1,2)$ and $G_c(1,2)$
using a Gaussian weight $e^{- \mr^2/2R_s^2}$ in the integrands,
extending the integration region to infinity. Specifically, we
have for $G_c(1,2)$ and $G_s(1,2)$
\bigskip
\begin{eqnarray}
G_c({\mk_1},{\mk_2}) &=& \frac{E_{1,2}}{(2 \pi)^3 } \int d^3r\, e^{-i(\mk_1 -
\mk_2){\bf \cdot}\mr} \Biggl\{ e^{-\mr^2/2R_s^2}\,\Biggl(
|c(1,2)|^2 n^*_{1,2}(x) \nonumber\\
&+& |s(-1,-2)|^2 \left[n^*_{-1,-2}(x) +1\right] \Biggr) + \left( 1
- e^{-\mr^2/2R_s^2} \right)\; n_{1,2}(x)  \Biggr\},
\label{gcs} \\
G_s({\mk_1},{\mk_2}) &=& \frac{E_{1,2}}{(2 \pi)^3 } \int d^3r \;
e^{-i(\mk_1 + \mk_2)\cdot{\mr}} e^{-\mr^2/2R_s^2} \Biggl(
s^*(-1,2) \, c(2,-1) \; n^*_{-1,2}(x) \nonumber \\
&+& c(1,-2) \, s^*(-2,1) \,\left[1 + n^*_{1,-2}(x)\right]\Biggr),
\label{gss}
\end{eqnarray}
where $n^*_{i,j}(x)$ means that the local distribution function is
to be evaluated with in-medium momenta, i.e. $n^*_{i,j}(x) \equiv
n(x,K^*_{i,j})$. The integral over the factor $\left( 1 -
e^{-{\mathbf r}^2/2R_s^2} \right)$ represents the integration over
the region where there is no mass shift, corresponding to the
region $V - V_s$. In this region, we have that the squeezing
factors become $c(i,j) \rightarrow 1$ and $s(i,j) \rightarrow 0$.

In order to proceed, we need the expressions for $n^*(i,j)$ and
$n(i,j)$. We consider their Boltzmann limit,
\be n(x,K_{i,j}) \approx  \exp{ \left[ - \frac{K^\mu_{i,j}
u_\mu(x) - \mu (x)}{T(x)}\right] }, \label{nij} \ee
and the same for $n^*_{i,j}(x)$ with $K_{i,j}(x)$ replaced by
$K^*_{i,j}(x)$. Considering that the chemical potential in the
model of Ref.~\cite{Csorgo:fg} can be written as $\mu(x)/T(x) =
\mu_0/T - \mr^2/2R^2$, and making use of Eq.~(\ref{Kstar}), it is
easy to show that
\bea n^*_{1,2}(\mr) &=& n^*_0~\exp\left\{{-\frac{{\mathbf
r}^2}{2R^2}- \frac{\left[({\mathbf k}_1+{\mathbf k}_2)/2- m
\langle u\rangle{\mathbf r}/R \right]^2}{2m_*T} - \frac{({\mathbf
k}_1-{\mathbf k}_2)^2}{8 m_* T}}\right\} \nn\\
&=& n^*_{-1,-2}(\mr) = n^*_{-1,2}(\mr) = n^*_{1,-2}(\mr),
\label{n12*}
\eea
where
\be n^*_0 = \exp \left(- \frac{m_* - \mu_0}{T}\right).
\label{n0}
\ee
This factor is proportional to the mean multiplicity, and can be
determined in principle from the absolute normalization of the
single particle spectra. The corresponding unstarred
$n_{i,j}(\mr)$ are obtained from $n^*_{i,j}(\mr)$ by replacing
$m_*$ by $m$ in Eq.~(\ref{n12*}).

When evaluating the spectra and the correlations from this model,
we realize that a mathematically equivalent problem has already
been considered in Ref.~\cite{Csorgo:fg}. By replacing $m
\rightarrow m_*$ and $t_0 \rightarrow R_G m_* / \langle u \rangle
m$ in the equations of Ref.~\cite{Csorgo:fg}, the results obtained
there can be directly transcribed here.

Due to the equality in Eq.~(\ref{nonrelK}), we see that the
accounting for the squeezing effects can be simplified for small
mass shifts $(m_* - m)/m \ll 1$, such that the squeezing parameter
can be written as
\begin{equation}
f(i,j,{\mathbf r})=\frac{1}{2}\log\left[\frac{K^{\mu}(i,j,x) \,
u_\mu (x)}{K^\nu_*(i,j,x) \, u_\nu(x)}\right] \approx
\frac{1}{2}\log\left(\frac{m}{m_*}\right). \label{e:rxvectk}
\end{equation}
The neglected terms are order of (kinetic energy/mass)$^2$
(masshift/mass)$^2$ and hence are of fourth order in small
quantities.  This limit is important, because in this case the
coordinate dependence enters the squeezing parameter $f$ only
through the possible position dependence of the mass-shift and so
the flow effects on the squeezing parameter can be neglected. In
principle, the position dependence of the mass shift can be
calculated from thermal field models in the local density
approximation. Therefore, in an approximation that the position
dependence of the in-medium mass is neglected, the $c(i,j)=c_0$ and
$s(i,j)=s_0$ factors can be removed from the integrands and all what
remains to be done are Fourier transforms of Gaussian functions.
The final result for $G_c$ and $G_s$ can be written as
\bea G_c({\mk_1},{\mk_2}) &=& \frac{E_{1,2}}{(2 \pi)^3} \Bigl[
n^*_0 \left(|c_0|^2 + |s_0|^2\right) I^c_{1,2}(R_s, R, T, m_*) +
|s_0|^2
I^c_{1,2}(R_s, \infty, \infty, m_*) \nn\\
&+& n_0 I^c_{1,2}(\infty, R, T, m) - n_0
I^c_{1,2}(R_s, R, T, m)\Bigr], \label{Gc12}\\
G_s({\mk_1},{\mk_2}) &=& \frac{E_{1,2}}{(2 \pi)^3} \,c_0|\,s_0|
\Bigl[ \, 2  \, n^*_0 \, I^s_{1,2} (R_s, R, T, m_*) + I^s_{1,2}
(R_s, \infty, \infty, m_*) \Bigr],\label{Gs12} \eea
where $I^c_{1,2}$ and $I^s_{1,2}$ are the resulting Fourier
integrals
\bea I^c_{1,2}(R_s, R, T, m_{(*)}) &=& (2\pi \rho^2_{(*)})^{3/2}
\exp\Biggl\{ - \frac{(\mk^2_1 +
\mk^2_2)}{4m_{(*)}T} \nn\\
&-& \frac{\rho^2_{(*)}}{2} \left[ (\mk_1 - \mk_2) + i
\frac{m\langle u\rangle (\mk_1 + \mk_2)}{2 m_{(*)} T R}
\right]^2\Biggr\},
\label{Ic12} \\
I^s_{1,2}(R_s, R, T, m_*) &=& (2\pi \rho^2_*)^{3/2}
\exp\Biggl\{-\frac{(\mk^2_1 + \mk^2_2)}{4 m_*T} -
\frac{\rho^2_*}{2}\Biggl[1 + i \frac{m\langle u\rangle}{2 m_* T R}
\Biggr]^2 \left(\mk_1 + \mk_2\right)^2\Biggr\}, \label{Is12} \eea
with
\be \frac{1}{\rho^2_{(*)}} \equiv \frac{1}{R^2_s} +
\frac{1}{R^2}\left( 1 + \frac{m^2 \langle u \rangle^2}{m_{(*)}
T}\right). \label{rho} \ee

Finally, we include finite-time emission effects in a schematic
way using for the invariant function $F(\tau)$, that appears in
the expression for $d^4\sigma^{\mu}(x)$, the following expression
\be
F(\tau) = \frac{\theta(\tau-\tau_0)}{\Delta t}
\;e^{-(\tau-\tau_0)/\Delta t},
\label{F}
\ee
where $\Delta t$ is a free parameter. This finishes the derivation
of all the ingredients needed to evaluate $C_2(\mk_1,\mk_2)$,
Eq.~(\ref{fullcorr}). In the next section we will present
numerical results.

\section{Numerical Results}
\label{sec:numeric}

We present numerical results for two situations regarding the
volumes over which mass modification occurs. In the first
situation the mass shift occurs over the entire volume of the
expanding system, i.e. $V_s = V$. This amounts to removing the
factor $e^{- \mr^2/2R_s^2}$ in Eqs.~(\ref{gcs}) and (\ref{gss})
or, equivalently, take $R_s \rightarrow  \infty$ in $I^c$ and
$I^s$. In the second situation, we consider $V_s < V$.
In order to comply with the non-relativistic nature of the
expansion model used in this paper, we present numerical results
for the Back-to-Back Correlations of a $\phi$ meson pair. In free
space, the $\phi$ meson mass is $m_\phi = 1020$~MeV.

\subsection{Bulk decay of a volume filled with mass-shifted quanta}

In this case, we suppose that the mass-shift occurs in the entire
volume of the system, for simplicity considered here as a Gaussian
with r.m.s. radius $R$. We will focus on the BBC correlation
function, whose generic expression consists of the first and third
terms on the right-hand-side of Eq. (\ref{fullcorr}). The detailed
expressions for the amplitudes are given in Appendix~\ref{appI}.
In what follows, we will concentrate on the value of momenta of
the participant pair that maximizes the BBC signal, i.e., the case
in which ${\mathbf k}_1  = - {\mathbf k}_2 = {\mathbf k}$. The BBC
correlation function can then be written as
\be C^{\sl max}_{BBC} ({\mk_1},{\mk_2}) = C_{BBC}({\mk},-{\mk}) =1
+ \frac{|G_s({\mk},-{\mk}) |^2}{G_c({\mk},{\mk})
G_c(-{\mk},-{\mk}) }= 1 + \frac{|G_s({\mk},-{\mk})
|^2}{\left[G_c({\mk},{\mk})\right]^2} \; .\label{BBCcorr} \ee
In the above equation, we have used the fact that the
single-inclusive distribution, $G_c({\mk},{\mk}) =
G_c(-{\mk},-{\mk})$ depends only on the absolute value of the
momentum, as can be seen in Eq.~(\ref{spectrum1Vf}). With the aid
of this equation, as well as of Eq. (\ref{Gs121Vf}), the
expression of the BBC correlation function is finally written as
\bea C^V_{BBC}({\mk},-{\mk}) &=& 1 + \big| c_0 \; s_0 \big|^2 \,
\Bigg| \frac{ 2 n^*_0 \; R_*^3 \exp ( - {\mathbf k}^2/2m_*T )  +
R^3 \; }{ n^*_0 R_*^3 \left( |c_0|^2 + |s_0|^2 \right)\exp ( -
{\mathbf k}^2/2m_*T_* ) + R^3 |s_0|^2
 }\Bigg|^2 , \label{BBCcorr1Vf}\eea
where
\beq
\frac{1}{R_*^2} = \frac{1}{R^2} \left( 1+ \frac{m^2\langle u
\rangle^2}{m_* T} \right), \hspace{1.0cm} T_* = T +
\frac{m^2}{m_*} \langle u\rangle^2 ,
\eeq
as given in Table~\ref{t1} of Appendix~\ref{appI}. The quantities
$R_*$ and $T_*$ are, respectively, the flow-modified radius and
the flow-modified temperature of the system  where the mass-shift
occurs in the entire volume.

\begin{figure}[htb]
\resizebox{18pc}{!}{\includegraphics{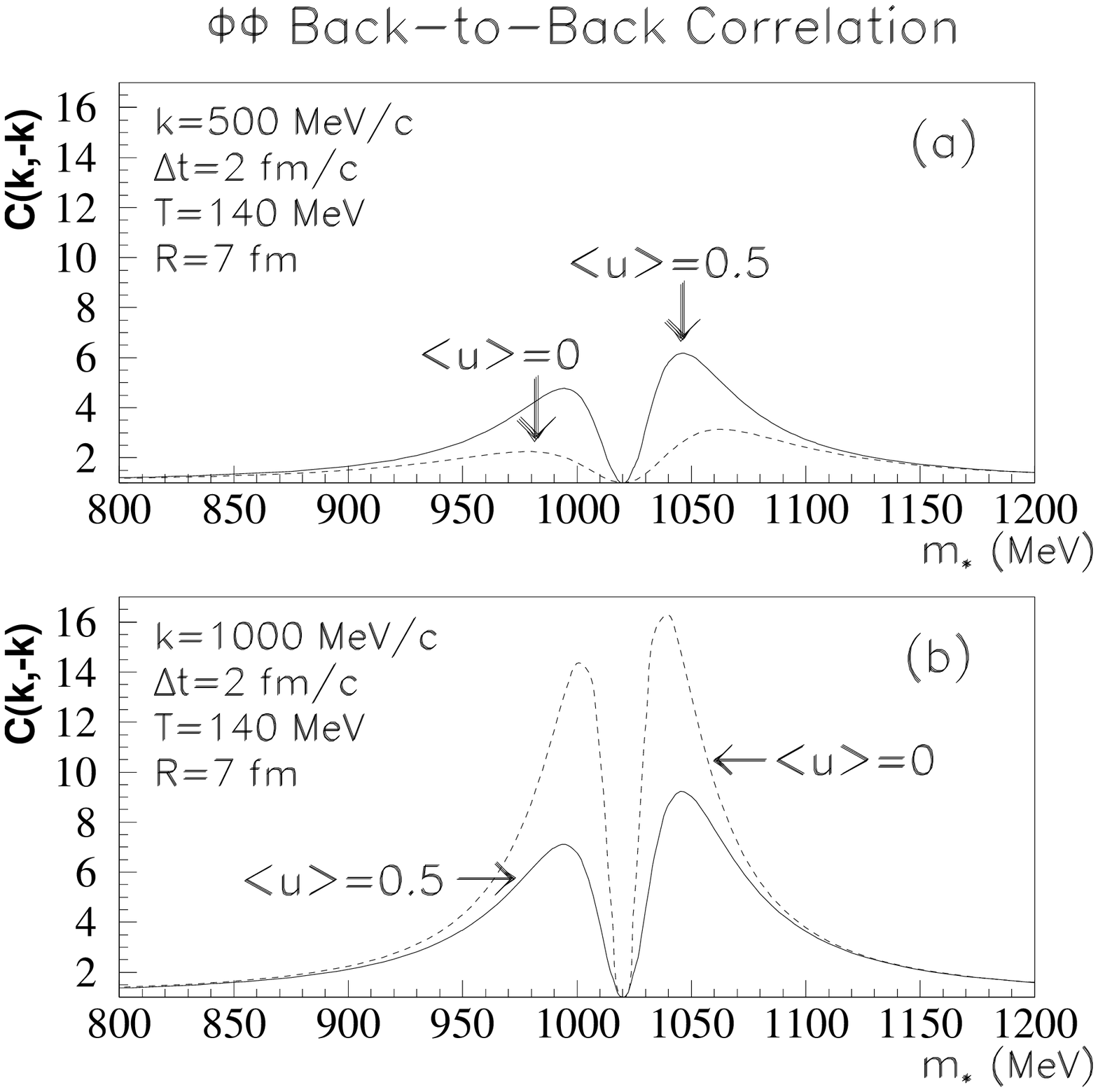}} \hspace{1cm}
\resizebox{18pc}{!}{\includegraphics{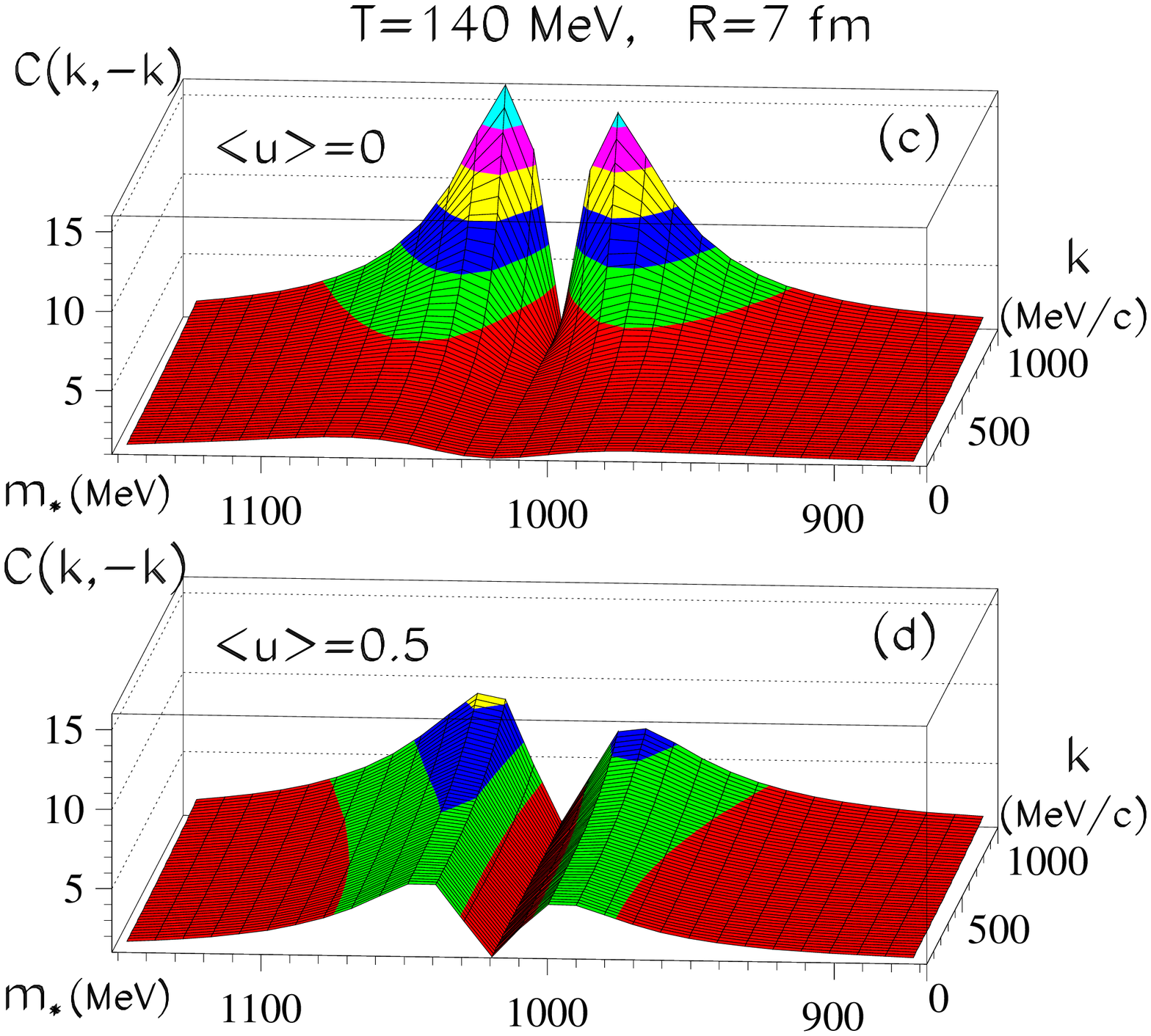}}
\caption{The Back-to-Back Correlation is shown as a function of
the shifted mass $m_*$ on the left panels, and as a function of
both $m_*$ and the momentum of each particle ($\mk_1=-\mk_2=\mk$)
on the right ones, for $R=7$ fm/c, $T=140$~MeV and $\Delta
t=2$~fm/c. The plots (a) and (b) illustrate better the behavior of
the BBC signal seen on the right panel, when $|\mk| = 500$~MeV/c
and for $|\mk| = 1000$~MeV/c, respectively. In both cases, the
dashed curve corresponds to $\langle u \rangle = 0$ and the solid
curve, to $\langle u \rangle = 0.5$. In (c), it is shown the 3-D
plot corresponding to the no flow case, with $\langle u \rangle
=0$, whereas in (d), a radial flow with $ v = \langle u \rangle
r/R = 0.5$ was considered. } \label{bbc1}
\end{figure}

In Fig.~\ref{bbc1} we present the results for $C^V_{BBC}$ for $T =
140$~MeV, $R = 7$~fm and a finite emission time $\Delta t =
2$~fm/c, for two different flow values. We clearly see in this
figure that the presence of a radial flow causes the BBC
correlation to be higher than the case with no flow in the low
momentum region and for the same values of $m_*$ and $\mk_1=-\mk_2
=\mk$, but it grows more slowly than in the no-flow case for
increasing values of $\mk$ and same $m_*$. We also see that for a
flow of $\langle u \rangle = 0.5$, the BBC signal increases for
values of the momenta $|\mk| \lesssim 1000$~MeV/c, but the no-flow
case surpasses the previous case for $|\mk| \gtrsim 1000$ MeV/c.
This conclusion is more easily achieved by looking into the right
panel of Fig. \ref{bbc1}. The inversion of the BBC behavior for
that value of $\mk$ roughly coincides with the limit of
applicability of our non-relativistic approximation. This result
is very stimulating, suggesting that we would have bigger chances
of observing the BBC effect in the lower $|\mk|$ side of the the
$BBC \times m_* \times |\mk|$ region.
\begin{figure}[h]
\resizebox{18pc}{!}{\includegraphics*{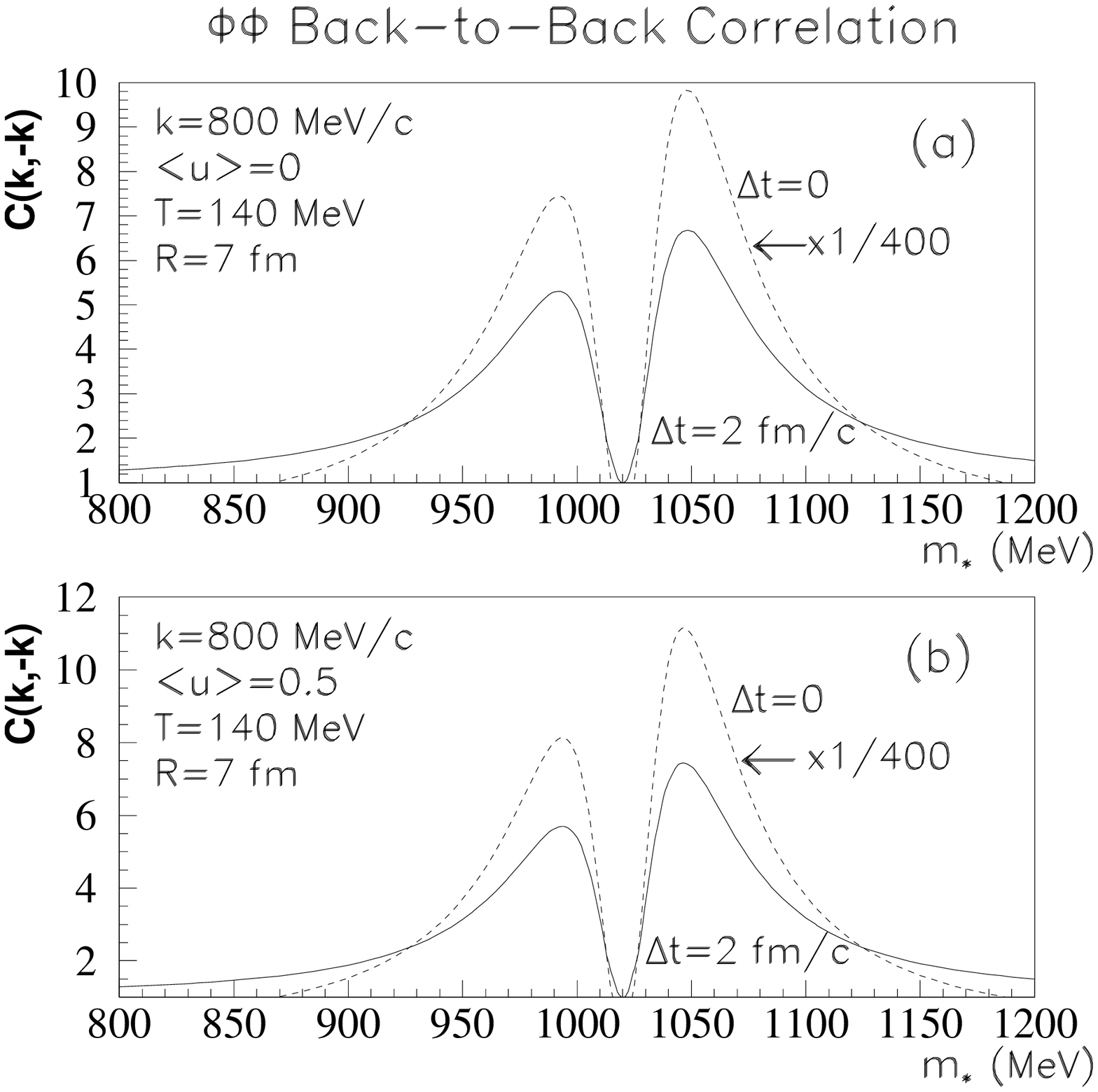}} \hspace{1cm}
\resizebox{18pc}{!}{\includegraphics*{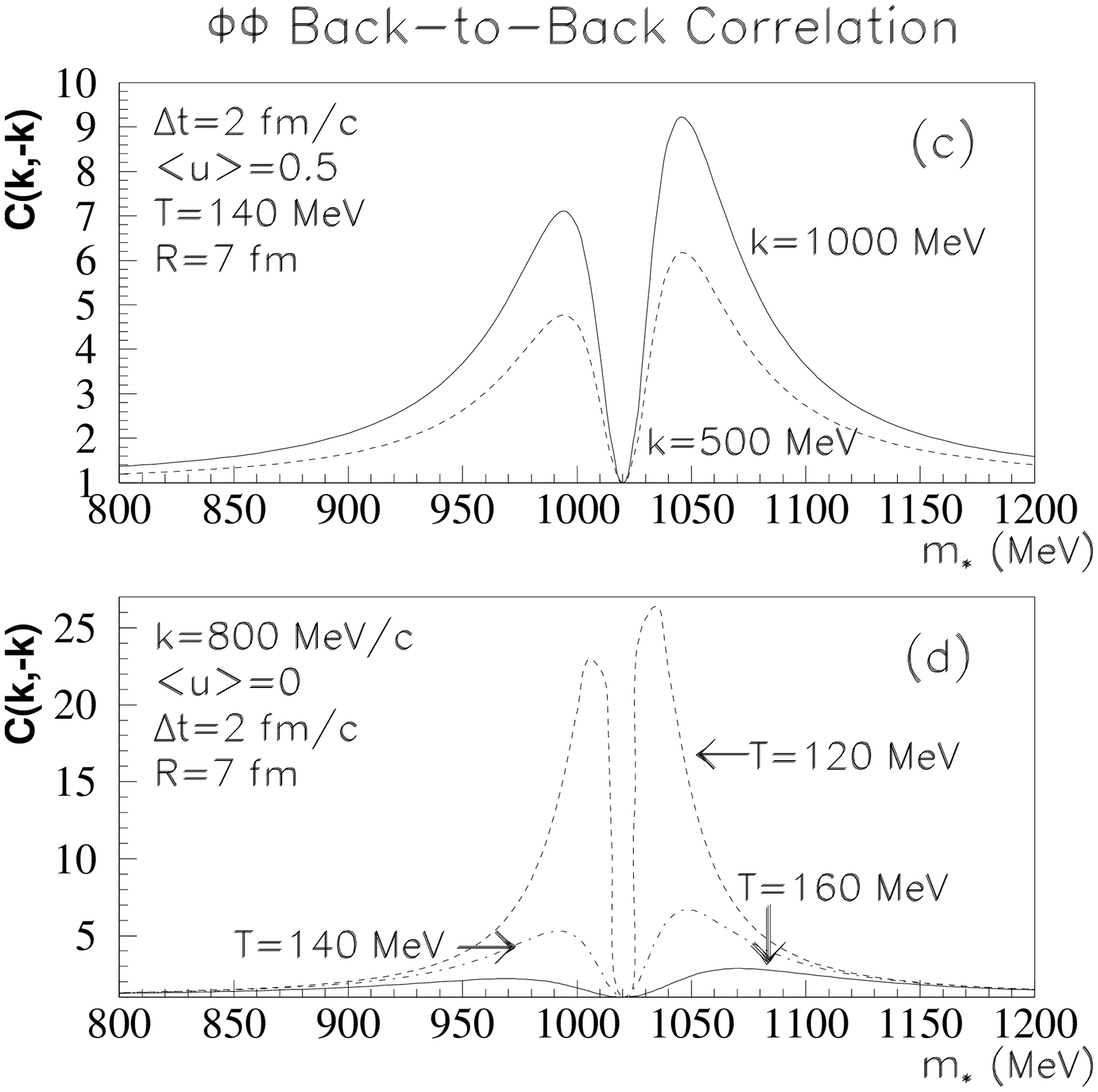}}
\caption{The effect of a finite emission interval on the Back to
Back correlation function, as compared to instant emission, is
illustrated in the left panel. The dashed curves have been reduced
by a factor of $400$, and the solid curves correspond to the
suppression by a finite emission duration, of about $\Delta t
\simeq 2$~fm/c. The plot in (a) shows this effect in the absence
of flow, already discussed in our previous paper. The plot  in (b)
shows the corresponding result when flow is included, with
$<\!\!u\!\!>\!=0.5$ (the other parameters adopted to produce the
curves are $R=7$ fm/c, $T=140$~MeV and $\Delta t=2$~fm/c. In part
(c), we fixed the other parameters and study the influence of
increasing values of $|\mk|$ (the dashed curve corresponding to
$|\mk|=500$~MeV/c and the solid one, to $|\mk|=500$~MeV/c).
Finally, the plot in (d) considers the variation of the BBC curve
for increasing values of $|\mk|$, as indicated by the arrows,
keeping all the other variables fixed to their values above. }
\label{bbc2}\end{figure}

In Fig.~\ref{bbc2} we present results for different combinations
of temperatures and flow velocities and emission times. We see
again that flow enhances the BBC correlation function for small
values of $|\mk|$. The effect of the temperature is such that the
BBC signal is stronger for lower values of $T$, and decreases the
signal for higher values of~$T$.

\subsection{Decay of a volume partly filled with mass-shifted quanta}

Here we consider the scenario in which the mass shift of the
bosons occurs in part of the system volume only. In this case, the
expression for the BBC correlation function is more complex, even
in the simpler limit of the maximal effect, i.e., when ${\mathbf
k}_1=-{\mathbf k}_2={\mathbf k}$. The expression of $C^{\sl
max}_{BBC}$ can again be obtained from Eq.~(\ref{BBCcorr}), but
this time we must replace the amplitudes by the expressions of
Eqs.~(\ref{Gs12-2Vf}) and (\ref{spectrum2Vf}) of
Appendix~\ref{appII}, in the limit that the particles are
back-to-back. In this case, we must use for the squeezing
parameters in the region where there is no medium modification
their appropriate limiting values
\beq \lim\limits_{m_* \rightarrow m} \left( \begin{array}{c}
  c_0 \\
  s_0 \\
\end{array}\right) \longrightarrow \left( \begin{array}{c}
  1 \\
  0 \\
\end{array} \right).
\eeq
With this, we obtain after a long but straightforward calculation
the expression
\bea
C^{2V}_{BBC}({\mk},-{\mk}) &=& 1+ \Big| c_0 \; s_0 \Big|^2 \; \Big|2 n^*_0 \;
\rho_*^3 \; \exp\!\!\left(
- \frac{{\mathbf k}^2}{2m_* T}\right)  + R_s^3 \Big|^2 \times \nn \\
&& \Biggl\{  n_0^* \;  \rho_*^3 \left( |c_0|^2+|s_0|^2 \right)
\exp{\left(-\frac{(R^2 +R_{_s}^2){\mathbf k}^2} {2m_* (R^2 T
+R_{_s}^2 T_*)}\right)}
+ R_s^3 |s_0|^2  \nn \\
&+& n_0 {\tilde{\cal R}}^3
\exp{\left(-\frac{{\mathbf k}^2}{2 m \tilde{\cal T}}\right)}  -
n_0 {\tilde{\rho}}^3
\exp{ \left(-\frac{(R^2 +R{_s}^2){\mathbf k}^2}
{2m (R^2 T +R{_s}^2 \tilde{\cal T})}\right)} \Biggr\}^{-2},\; \nn \\
\label{BBCcorr2Vf}\eea
where the parameters $\tilde{\cal T} = T + m \langle u\rangle^2$,
$\tilde{\cal R}^{-2}=R^{-2} \left( 1+ m\langle u \rangle^2/T
\right)$, $\tilde{\rho}^{-2}=\tilde{\cal R}^{-2} + R_s^{-2}$,
$\rho_*^{-2}=R^{-2} \left( 1 + m^2\langle u \rangle^2 /m_* T
\right)+ R_s^{-2}$ are given in Table~\ref{t2} of
Appendix~\ref{appII}. The two are, respectively, the flow-modified
temperature and the flow-modified radius in the region of no
mass-shift. On the other hand, $\tilde{\rho}$ and $\rho_*$ are
effective radius parameters corresponding to the no mass-shift
region and the region where the mass-shift occurred, respectively.
We see that they are functions of the flow parameter $\langle u
\rangle$ and the parameter $R_s$, which corresponds to the radius
of the mass-shift region. The parameter $T_* = T + m^2 \langle
u\rangle^2 / m_*$, is the same as before, also written in
Table~\ref{t1} of Appendix~\ref{appI}.

\begin{figure}[h]
\resizebox{18pc}{!}{\includegraphics*{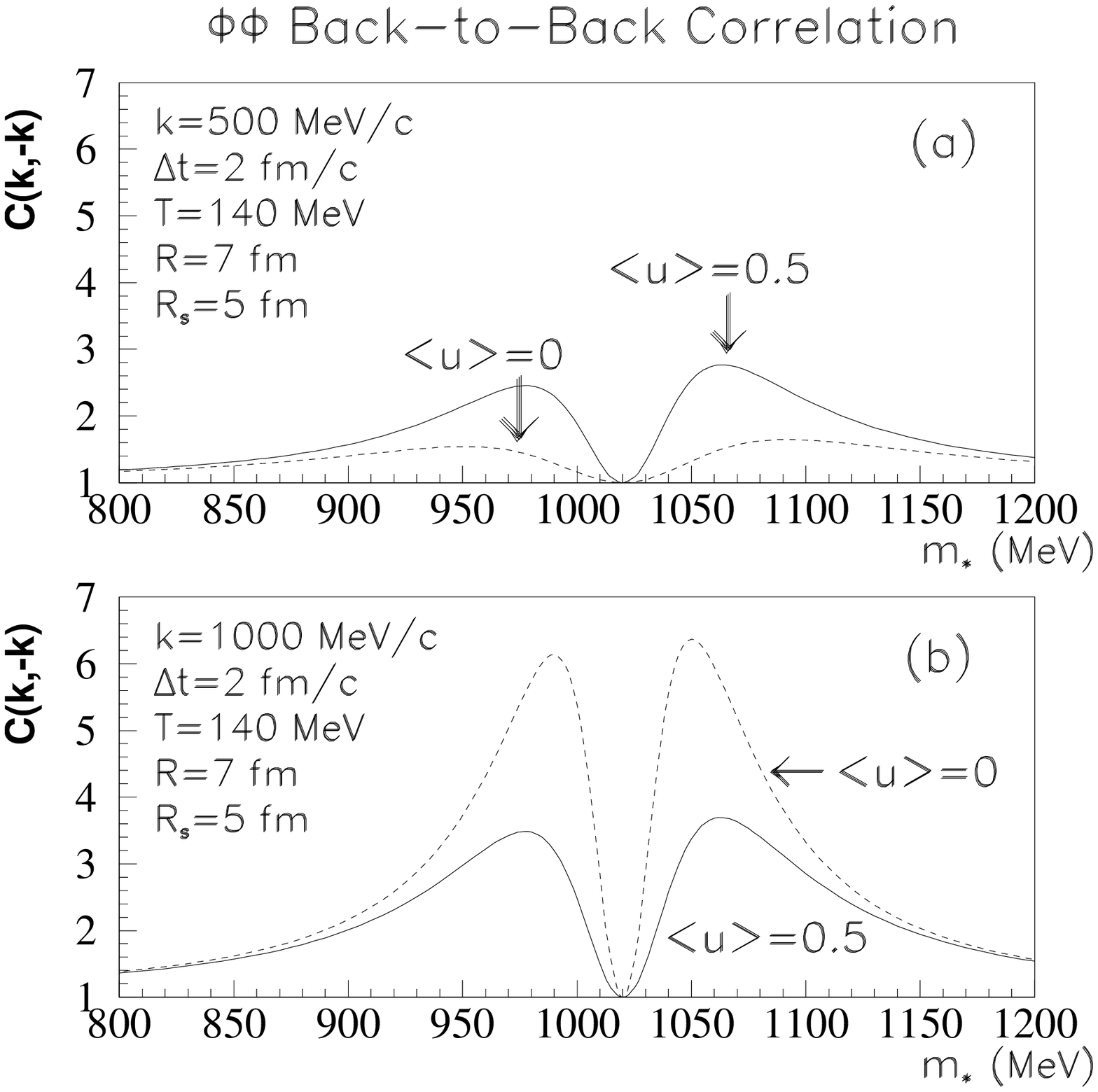}} \hspace{1cm}
\resizebox{18pc}{!}{\includegraphics{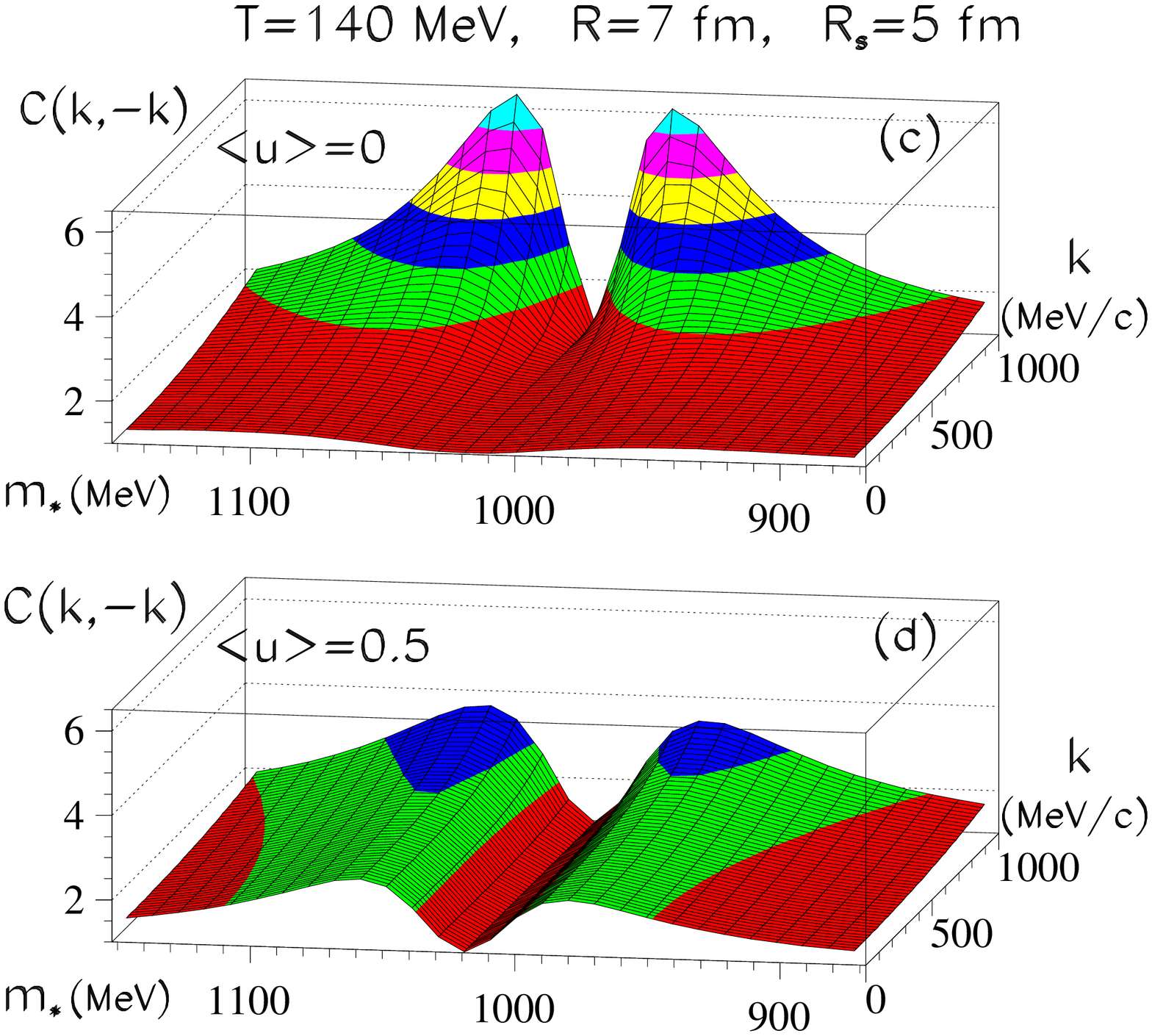}}
\caption{The plots in this panel are similar to the ones on
Fig.~\ref{bbc1}. The main difference is that here we assumed that
the mass-shift occurred only in a smaller part of the system
volume. Here, the back-to-back correlation is shown as a function
of the shifted mass $m_*$ on the right panel, and as function of
both $m_*$ and the momentum of each particle ($\mk_1=-\mk_2=\mk$),
on the left ones, for $R=7$~fm/c, $R_s=5$ fm/c, $T=140$~MeV and
$\Delta t=2$~fm/c. The plots in parts (a) and (b) illustrate
better the behavior of the BBC signal seen in parts (c) and (d),
for $|\mk| = 800$~MeV/c and for $|\mk| = 1000$~MeV/c. In both
cases, the dashed curve corresponds to $\langle u \rangle = 0$ and
the solid curve, to $\langle u \rangle =0.5$. In the 3-D plot in
(c), no flow ($\langle u \rangle =0$) was considered, whereas a
radial flow with $v = \langle u \rangle r/R= 0.5$ was included in
the (d) plot. } \label{bbc3}\end{figure}

Similarly to Eq. (\ref{BBCcorr1Vf}), the above Eq.
(\ref{BBCcorr2Vf}) has also very interesting limiting cases. The
first one is the case of vanishing squeezing, $m_* \rightarrow m$,
which implies that $|s_0 | \rightarrow 0$ and $|c_0| \rightarrow
1$, and the squeezed Back-to-Back Correlations vanish.

The large momentum limit is also very interesting. In this case,
the exponential, thermal contributions disappear, and the
surviving terms come from the decay of the squeezed vacuum to the
asymptotic quanta. Both the numerator and the denominator of Eq.
(\ref{BBCcorr2Vf}) will be proportional to the square of the
squeezed volume, hence Eq. (\ref{BBCcorr2Vf}) and Eq.
(\ref{BBCcorr1Vf}) will be reduced to a form similar to
$$
    C(k,-k) = 1 + |c_0/s_0|^2
$$
which has no upper limit, and diverges for small but non-vanishing
amount of squeezing, where $|c_0| \rightarrow 1$ and $|s_0|
\rightarrow 0$. This property, the unlimited strength  of the
squeezed BBC-s -- even if the mass modification does not happen in
the whole volume -- makes it worthwhile to look for these effects
experimentally as signals of in-medium mass modifications. Again,
for large in medium mass modifications and large momenta, the
strength of the squeezed BBC-s will be similar to that of the HBT
effect:
$$
    C(k,-k) \rightarrow  2
$$
if $|c_0|, |s_0| \rightarrow \infty $, as in this limit,
$|c_0|/|s_0| \rightarrow 1$.

The single particle spectra also behaves interestingly in these
limits, which is discussed in Appendices~\ref{appI}
and~\ref{appII}.

In Fig. \ref{bbc3} we show the BBC correlation corresponding to
the hypothesis that the mass-shift occurred only in a smaller part
of the system volume. We see a very close similarity to the
results corresponding to the mass-shift occurring in the entire
system volume, shown in Fig. \ref{bbc1}. The major difference
between the two of them is that in Fig. \ref{bbc3} the correlation
signal is lower, as expected, since a the mass-shift occurred in a
smaller volume in this case.

\vskip1cm

\section{Conclusions and Future Perspectives}
\label{sec:concl}

In this paper we discussed the effects of the system expansion and
flow on the back-to-back correlation, also limiting the system to
a more realistic finite size. For simplicity, we restricted our
analysis to the non-relativistic domain. In our study, we have
also considered that the flow effects on the squeezing parameter
were negligible. For simplicity, we have also assumed a 3-D
Gaussian profile for the system. We showed here the effects of the
decoupling temperature on the BBC signal, fixing all other
parameters, as in the bottom right plot of Fig. \ref{bbc2}. In
this case, we observed that the BBC signal survives stronger if
the decoupling temperature is lower. Fixing $T$ and the other
parameters, we also showed the effect of increasing momentum on
the signal survival. More striking, we showed in Figures
\ref{bbc1} and \ref{bbc3} the conclusion about the best region for
looking into the BBC effect in the $m_* \times \vert \mk \vert$
plane: the search for the signal is more pronounced in the small
$|\mk|$ region, when we are to take into account the system
expansion and the presence of moderate to strong flow. For higher
values of $\vert \mk \vert$, the BBC signal would be more
pronounced if the system flow could be neglected. In any case,
what remains as the most encouraging point coming out of our
present study is that the BBC seems to survive with measurable
intensity in the more realistic situation of finite size systems
subjected to hydrodynamical expansion and consequent flow.

\section*{Acknowledgments}
We are deeply grateful to Prof. Mikl\'os Gyulassy for estimulating
discussions and to Prof. Roy Glauber for inspiring conversations.
This research has been supported in part by CNPq, FAPESP grants
00/04422-7, 04/10619-9, 02/11344-8, 99/08544-0, by the Hungarian
OTKA T038406, T043514 and T049466, by a Hungarian - US MTA - OTKA
- NSF grant and by the NATO PST CLG.980086 grant.

%%%%%%%%%%%%%%%%%%%%%%%%%%%%%%%%%%%%%%%%%%%%%%%%%%%%%%%%%%%%%%%%%%%%%%%%%%%%%%%%%%%%
\vskip 1cm

\appendix

\section{Mass-shifted in the entire volume}
\label{appI}

It turns out that the formalism can be presented in the simplest
manner if we assume, that the whole thermalized medium is filled
with mass-shifted quanta, and that the whole medium decays
suddenly to asymptotic quanta. This case is the subject of this
appendix. It is also possible, that e.g. due to density
inhomogeneities, the volume where the mass-shift is non-zero is
different (smaller) than the totality of the volume filled out by
thermalized quanta. This case will be investigated in the next
appendix.

For the case of non-relativistic hydrodynamics, assuming for the
sake of simplicity a sudden freeze-out, $\int d^4 \Sigma_\mu
K^\mu_{ij} = \int d^3 \mr \int dt \delta(t - t_0) E_{ij} = E_{ij}
\int d^3\mr$, the chaotic and the squeezed amplitudes are easily
obtained from the ones previously derived in Eq. (\ref{Gc12}) and
(\ref{Gs12}) by taking the limit $R_s \rightarrow \infty$ in those
equations as well as in all the others that immediately follow
them, i.e., Eq. (\ref{Ic12}), (\ref{Is12}), and (\ref{rho}). The
intuitive way to understand this limit is to consider the
mass-shift region as extended so as to include the entire volume
of the system, by simply taking the limit $R_s \rightarrow
\infty$. The volume of the system, however, will be still
delimited by the Gaussian profile with  rms $R$. In this limit,
the last two terms in Eq.(\ref{Gc12}) exactly cancel, since
\beq \lim\limits_{R_s \rightarrow \infty} I^c_{1,2}(R_s, R,
\langle u \rangle, m) = I^c_{1,2}(\infty, R, \langle u \rangle,
m). \eeq
Consequently, the effective squeezing region becomes $R_s
\rightarrow R$.

Before writing the resulting expression for $G_c(1,2)$, $G_s(1,2)$
and $G_c(i,i)$, it is usefull to define two parameters in terms of
which we can write those expressions, i.e., the flow-modified
temperature $T_*$ and the flow-modified radius of the single
volume case
\beq R_* = \lim\limits_{R_s \rightarrow \infty}
\rho(R_s,R,m_{(*)}) \eeq
as given in Eq. (\ref{rho}). We can then write the chaotic
amplitude, $G_c(\mk_1,\mk_2)$ as
\bea G^{1V}_c(\mk_1,\mk_2) &=& \lim\limits_{R_s \rightarrow
\infty} \left\{ \frac{E_{1,2}}{(2 \pi)^3} \Bigr[ n^*_0
\left(|c_0|^2 + |s_0|^2\right) I^c_{1,2}(R_s, R, T, m_*) +
|s_0|^2 I^c_{1,2}(R_s, R, \infty, m_*) \Bigr]  \right\}  \nn\\
& = & \frac{E_{1,2}}{(2 \pi)^3} |s_0|^2 (2\pi R^2)^{3/2}
\exp\left[ - \frac{R^2}{2} (\mk_1 - \mk_2)^2\right] \nn
\\
&+& \frac{E_{1,2}}{(2 \pi)^3} n^*_0 \left( |c_0|^2 + |s_0|^2
\right)  \nn \\
&\times& (2\pi R_*^2)^{3/2} \exp\left\{ -\frac{(\mk^2_1 +
\mk^2_2)}{4m_*T} - \frac{R_*^2}{2} \left[ (\mk_1 - \mk_2) + i
\frac{m\langle u\rangle (\mk_1 + \mk_2)}{2 m_* T R} \right]^2
\right\} . \label{Gc121V}\eea
We can rearrange the above terms in a more compact form, and
explicitly writing in terms of the variables defined in
Table~\ref{t1} (from which we can see that $R^2_*/R T = R/T_*$) we
have
\bea G^{1V}_c(\mk_1,\mk_2) &=& \frac{E_{1,2}}{(2 \pi)^{3/2}} R^3
|s_0|^2 \exp\left[ - \frac{R^2}{2} (\mk_1 - \mk_2)^2\right] \nn
\\
&+& \frac{E_{1,2}n^*_0}{(2 \pi)^3} \; (|c_0|^2 + |s_0|^2) \; (2\pi
R^2_*)^{3/2} \exp\left[-\frac{({\mathbf k}_1+{\mathbf k}_2)^2}{8
m_* T_*}\right] \nn\\
&\times & \exp\left[-\frac{i m \langle u\rangle R}{2 m_*
T_*}({\mathbf k}_1^2- {\mathbf k}_2^2)
-\left(\frac{1}{8m_*T}+\frac{R_*^2}{2}\right) ({\mathbf
k}_1-{\mathbf k}_2)^2 \right] .
\label{Gc121Vf}\eea

\begin{table}[ht]
\caption{\label{t1} Parameters used in the single volume case.}
\begin{ruledtabular}
\begin{tabular}{c|l|c}
Parameter& Relation to other parameters & Integral results where they appear  \\
\hline
$T_*$ &$T_*=T+\frac{m^2}{m_*} \langle u\rangle^2$ &  $I^c_{i,j}(\infty, R, T, m_*)$ \\
\cline{1-2} $R_*$ &  $R_*^{-2}=R^{-2} \left( 1+ \frac{m^2\langle u
\rangle^2}{m_* T} \right)$  & $I^s_{i,j}(\infty, R, T, m_*)$
\\
\end{tabular}
\end{ruledtabular}
\end{table}

\bigskip\smallskip

The coherent amplitude $G_s(\mk_1,\mk_2)$ can be written as
\bea
G^{1V}_s(\mk_1,\mk2) &=&  \lim\limits_{R_s \rightarrow
\infty} \left\{ \frac{E_{1,2}}{(2 \pi)^3} \, c_0|s_0| \Bigl[ 2 \,
n^*_0 I^s_{1,2}(R_s, R, T, m_*) +
I^s_{1,2}(R_s, R, \infty, m_*) \Bigr] \right\}\nn \\
&=& \frac{E_{1,2}}{(2 \pi)^3} c_0|s_0| (2\pi R^2)^{3/2} \exp\left[
- \frac{R^2}{2} (\mk_1 + \mk_2)^2\right] +
\frac{E_{1,2}n^*_0}{(2 \pi)^3}  \, 2 c_0|s_0| \nn\\
\nn \\
&\times &  (2\pi R_*^2)^{3/2} \, \exp\left[ -\frac{(\mk^2_1 +
\mk^2_2)}{4 m_*T} - \frac{R_*^2}{2}\left(1 + i \frac{ m \langle u
\rangle}{2 m_* T R}\right)^2 \left(\mk_1 + \mk_2\right)^2\right] .
\label{Gs121V}
\eea

Similarly to what was done before, can also rewrite the expression
for $G_s(\mk_1,\mk_2)$ explicitly in terms of the variables
defined in Table~\ref{t1}, leading to
\bea
G^{1V}_s(\mk_1,\mk_2) &=& \frac{E_{1,2}}{(2 \pi)^3} c_0|s_0|
(2\pi R^2)^{3/2} \exp\left[ - \frac{R^2}{2} (\mk_1 +
\mk_2)^2\right] \nn \\
&+& \frac{E_{1,2} n^*_0}{(2 \pi)^{3/2}} \; (2\pi R^2_*)^{3/2} \,
(2 \, c_0|s_0|)\, \exp\left[ -\frac{({\mathbf k}_1 - {\mathbf
k}_2)^2}{8m_*T}\right] \nn \\
&\times& \exp\left[-\frac{i m \langle u\rangle R}{2 m_*
T_*}({\mathbf k}_1 + {\mathbf k}_2)^2 - \left(\frac{1}{8m_*T_*} +
\frac{R_*^2}{2}\right) ({\mathbf k}_1+{\mathbf k}_2)^2 \right] .
\label{Gs121Vf} \eea
Also, the single-particle distribution,  the amplitude appearing
in the denominator of both the BBC and the HBT correlation
functions can be written as
\bea N^{1V}_1({\bf k}_i) &=&  G^{1V}_c(\mk_i,\mk_i) \nn \\
&=& \lim\limits_{R_s \rightarrow \infty} \frac{E_{i,i}}{(2 \pi)^3}
\Bigr[ n^*_0\left(|c_0|^2 + |s_0|^2\right) I^c_{i,i}(R_s, R, T,
m_*) + |s_0|^2 I^c_{i,i}(R_s, R, \infty, m_*) \Bigr]  \nn\\
&=& \frac{E_{i,i}}{(2 \pi)^3} |s_0|^2 (2\pi R^2)^{3/2} \nn
\\
&+& \frac{E_{i,i}  n^*_0  }{(2 \pi)^3} \left( |c_0|^2 + |s_0|^2
\right) (2\pi R_*^2)^{3/2} \exp\left[ -\frac{\mk^2_i}{2m_*T} +
\frac{R_*^2}{2} \left( \frac{m\langle u\rangle \mk_i }{m_* T R}
\right)^2 \right] \nn\\
\label{spectrum1V}
\eea
Analogously, we can rewrite $G_c(\mk_i,\mk_i)$ explicitly in terms of
$T_*$ and $R_*$, as
\bea N^{1V}_1({\bf k}_i) &=&  G^{1V}_c(\mk_i,\mk_i) \nn \\
&=& \frac{E_{i,i}}{(2 \pi)^3} |s_0|^2 (2\pi R^2)^{3/2}  +
\frac{E_{i,i}n^*_0}{(2 \pi)^{3/2}}\left( |c_0|^2 + |s_0|^2 \right)
(2\pi R^2_*)^{3/2} \exp\left( -\frac{\mk^2_i}{2m_*T_*} \right).
\label{spectrum1Vf} \eea

Let us investigate the vanishing squeezing and the large momentum
limits of the single particle spectra, similarly to the analysis
of the correlation functions as was done after Eq.
(\ref{BBCcorr1Vf}).

In case of vanishing squeezing, $m_* \rightarrow m$, $|s_0 |
\rightarrow 0$ and $|c_0| \rightarrow 1$, hence we spectra will
contain a thermal and  a flow contribution, and we  recover the
results of Ref. \cite{Csorgo:fg}. In the large momentum limit, for
non-vanishing squeezing, rather surprisingly the single particle
spectra becomes a constant. This corresponds to the decay of a
modified vacuum with a fixed volume, described by the first term
of (\ref{spectrum1Vf}). This is the direct consequence of our
neglecting for the present purposes the position dependence of the
in-medium mass modification. Also, this result implies that the
squeezing mechanism not only makes strong signals in the
back-to-back correlations, but there is also an interesting signal
for squeezing in the single particle spectra.

\section{Mass-shifted in partial volume}
\label{appII}

If the mass-shift occurs only in a certain portion of volume $V_s$
of the whole system $V (> V_s)$, the expressions for the
amplitudes contain other terms besides the ones discussed in the
Appendix~\ref{appI}. Again, in order to avoid too much clutter it
is useful to define appropriate flow-modified variables. However,
in this case, we will need to define two sets of such parameters
as flow-modified radii and temperatures, one set corresponding to
the region where there is no mass-shift, which we will denote by
$\tilde{\cal R}$, $\tilde{\rho}$, and $\tilde{\cal T}$, and
another for the inside of the mass-shifted region, denoted by
$R_{*}$ and $T_*$, this last one, naturally, being the same as
defined in Table~\ref{t1}.

\bigskip

\begin{table}[ht]
\caption{\label{t2} Parameters used in the two-volume case.}
\begin{ruledtabular}
\begin{tabular}{c|l|c}
Parameter    & Relation to other parameters & Integral results where they appear \\
\hline $\tilde{\cal T}$ & $\tilde{\cal T} = T + m \langle
u\rangle^2$ & $I^c_{i,j}(\infty, R, T, m)$ \& $I_c(R_s, R, T, m)$  \\
\hline $\tilde{\cal R}$ & $\tilde{\cal R}^{-2}=R^{-2} \left( 1+
\frac{m}{T}\langle u \rangle^2 \right)$ &  $I^c_{i,j}(\infty, R, T, m)$ \\
\hline
$\tilde{\rho}$ & $\tilde{\rho}^{-2}=\tilde{\cal R}^{-2} +
R_s^{-2}$ & $I^c_{i,j}(R_s, R, T, m)$ \\
\hline
$T_*$ &  $T_*=T+\frac{m^2}{m_*} \langle u\rangle^2$ &  $I^c_{i,j}(R_s, R, T, m_*)$  \\
\cline{1-2} $\rho_*$ &  $\rho_*^{-2}=R^{-2} \left( 1+
\frac{m^2\langle u \rangle^2}{m_* T} \right)+R_s^{-2}$  &
$I^s_{i,j}(R_s, R, T, m_*)$ \\
\end{tabular}
\end{ruledtabular}
\end{table}

In the case where the mass-shift occurs in a small portion of the
system volume, $V_s < V$, the chaotic amplitude is given by Eq. (\ref{Gc12})
and (\ref{Ic12}), i.e.,
\bea
G^{2V}_c(1,2) &=& \frac{E_{1,2}}{(2 \pi)^3} \Bigl[ n^*_0 \left(|c_0|^2 +
|s_0|^2\right) I^c_{1,2}(R_s, R, T, m_*) + |s_0|^2
I^c_{1,2}(R_s, \infty, \infty, m_*) \nn\\
&+& n_0 I^c_{1,2}(\infty, R, T, m) - n_0 I^c_{1,2}(R_s, R, T,
m)\Bigr] . \label{Gc12-2V} \eea
Working out each of the integrals above separately, we have
\bea I^c_{1,2}(R_s, R, T, m_*) &=& (2\pi \rho_*^2)^{3/2}
\exp\left\{ - \frac{(\mk^2_1 + \mk^2_2)}{4m_* T} -
\frac{\rho_*^2}{2} \left[ (\mk_1 - \mk_2) + i \frac{m\langle
u\rangle (\mk_1 + \mk_2)}{2 m_* T R}
\right]^2\right\} \nn\\
&=& (2\pi \rho_*^2)^{3/2} \exp \left[-i \frac{m \langle u\rangle
\rho_*^2 ({\mathbf k}_1^2-{\mathbf k}_2^2)}{2 m_* RT} \right] \nn
\\
&\times& \exp{\left[ -\left(\frac{1}{8 m_*T} +
\frac{\rho_*^2}{2}\right) ({\mathbf k}_1-{\mathbf k}_2)^2
-\frac{(R^2+R_{_s}^2)({\mathbf k}_1+{\mathbf k}_2)^2} {8 m_* (R^2
T +R_{_s}^2 T_*)} \right]} \label{Ic12-3} , \\
I^c_{1,2}(R_s, \infty, \infty, m_*) &=& (2\pi R_s^2)^{3/2}
\exp\left[ - \frac{R_s^2}{2} (\mk_1 - \mk_2) ^2\right] ,
\label{Ic12-4} \\
I^c_{1,2}(\infty, R, T, m) &=& (2\pi \tilde{\cal R}^2)^{3/2}
\exp\left\{ - \frac{(\mk^2_1 + \mk^2_2)}{4 m T} -
\frac{\tilde{\cal R}^2}{2} \left[ (\mk_1 - \mk_2) + i
\frac{m\langle u\rangle (\mk_1 + \mk_2)}{2 m T R}
\right]^2\right\} \nonumber\\
&=& (2\pi {\tilde{\cal R}}^2)^{3/2} \exp\left[ -i \frac{\langle
u\rangle {\tilde{\cal R}}^2({\mathbf k}_1^2-{\mathbf k}_2^2)}
{2RT} \right] \nn \\
&\times& \exp{\left[ -\left(\frac{1}{8mT} + \frac{{\tilde{\cal
R}}^2}{2}\right) ({\mathbf k}_1-{\mathbf k}_2)^2 -\frac{({\mathbf
k}_1+{\mathbf k}_2)^2}{8 m \tilde{\cal T}} \right]} ,
\label{Ic12-1}\\
I^c_{1,2}(R_s, R, T, m) &=& (2\pi \tilde{\rho}^2)^{3/2}
\exp\left\{ - \frac{(\mk^2_1 + \mk^2_2)}{4 m T} -
\frac{\tilde{\rho}^2}{2} \left[ (\mk_1 - \mk_2) + i \frac{m\langle
u\rangle (\mk_1 + \mk_2)}{2 m T R}
\right]^2\right\} \nonumber\\
&=&(2\pi {\tilde{\rho}}^2)^{3/2} \exp\left[-i \frac{\langle
u\rangle {\tilde{\rho}}^2 ({\mathbf k}_1^2-{\mathbf k}_2^2)}{2 R
T} \right] \nn \\
&\times& \exp{\left[ -\left(\frac{1}{8mT} +
\frac{{\tilde{\rho}}^2}{2} \right) ({\mathbf k}_1-{\mathbf k}_2)^2
-\frac{(R^2+R_{_s}^2)({\mathbf k}_1+{\mathbf k}_2)^2} {8 m (R^2 T
+R_{_s}^2 \tilde{\cal T})} \right]}. \label{Ic12-2}
\eea
\\
Substituting these into Eq. (\ref{Gc12-2V}), the complete
expression for the chaotic amplitude can finally be written as
\begin{eqnarray}
G^{2V}_c(1,2) &=& n_0^* \; \rho_*^3  \left[ |c_0|^2 + |s_0|^2
\right] \exp\left[-i \frac{m \langle u\rangle \rho_*^2 ({\mathbf
k}_1^2-{\mathbf k}_2^2)}{2 m_* RT} \right]
\nonumber\\
&\times & \exp \left[ -\left(\frac{1}{8 m_*T} +
\frac{\rho_*^2}{2}\right) ({\mathbf k}_1-{\mathbf k}_2)^2
-\frac{(R^2+R_{_s}^2)({\mathbf k}_1+{\mathbf k}_2)^2} {8 m_* (R^2
T + R_{_s}^2 T_*)} \right] \nonumber\\
&+& \frac{E_{1,2}}{(2 \pi)^{3/2}} R_{s}^3 |s_0|^2 \exp
\left[-\frac{R_{s}^2}{2} ({\mathbf k}_1-{\mathbf k}_2)^2 \right]
\nonumber\\
&+& \frac{E_{1,2} n_0}{(2 \pi)^3} \Biggl\{  (2\pi {\tilde{\cal
R}}^2)^{3/2} \exp \left[-i \frac{\langle u\rangle {\tilde{\cal
R}}^2 ({\mathbf k}_1^2-{\mathbf k}_2^2)}{2 R T} \right]
\nonumber\\
&\times & \exp{\left[ -\left(\frac{1}{8mT} +
\frac{{\tilde{\cal R}}^2}{2} \right)
({\mathbf k}_1-{\mathbf k}_2)^2
-\frac{({\mathbf k}_1+{\mathbf k}_2)^2}{8 m \tilde{\cal T}} \right]}
\nonumber\\
&-& (2\pi {\tilde{\rho}}^2)^{3/2} \exp \left[-i \frac{\langle
u\rangle {\tilde{\rho}}^2 ({\mathbf k}_1^2-{\mathbf k}_2^2)}{2 R
T} \right] \nonumber\\
&\times & \exp \left[ -\left(\frac{1}{8mT} +
\frac{{\tilde{\rho}}^2}{2} \right) ({\mathbf k}_1-{\mathbf k}_2)^2
-\frac{(R^2+R_{_s}^2)({\mathbf k}_1+{\mathbf k}_2)^2} {8 m (R^2 T
+R_{_s}^2 \tilde{\cal T})} \right] \Biggr\} \label{Gc12-2Vf}
\end{eqnarray}

\bigskip
The coherent amplitude is given by Eqs.~(\ref{Gs12}) and
(\ref{Is12}), as
\beq G^{2V}_s(\mk_1,\mk2) = \frac{E_{1,2}}{(2 \pi)^3} \,c_0|\,s_0|
\Bigl[ \, 2  \, n^*_0 \, I^s_{1,2}(R_s, R, T, m_*) +
I^s_{1,2}(R_s, \infty, \infty, m_*) \Bigr],\label{Gs12-2V} \eeq
where
\bea
I^s_{1,2}(R_s, R, T, m_*) &=& (2\pi \rho_*^2)^{3/2}
\exp\left\{-\frac{(\mk^2_1 + \mk^2_2)}{4 m_*T} -
\frac{\rho_*^2}{2}\left[1 + i \frac{m\langle u\rangle}{2 m_* T R}
\right]^2\left(\mk_1 + \mk_2\right)^2\right\}
\nonumber\\
&=& (2\pi \rho_*^2)^{3/2} \; \exp{\left[-i \frac{m \langle
u\rangle \rho_*^2 ({\mathbf k}_1+{\mathbf k}_2)^2}{2 m_* RT}
\right]} \nonumber\\
&\times &\exp{\left[ - \frac{({\mathbf k}_1-{\mathbf k}_2)^2}{8
m_*T} - \frac{\rho_*^2}{2} ({\mathbf k}_1+{\mathbf k}_2)^2
-\frac{(R^2+R_{_s}^2)({\mathbf k}_1+{\mathbf k}_2)^2} {8 m_* (R^2
T +R_{_s}^2 T_*)} \right]} , \label{Is12-1} \\
I^s_{1,2}(R_s, \infty, \infty, m_*) &=& (2\pi R_{s}^2)^{3/2}
\exp\left[-\frac{R_{s}^2}{2}\left(\mk_1 + \mk_2\right)^2\right] .
\label{Is12-2}\eea
By substituting the above two terms, Eq. (\ref{Is12-1}) and
(\ref{Is12-2}) into Eq.~(\ref{Gs12-2V}), we get the final form of
the coherent amplitude
\begin{eqnarray}
G^{2V}_s(1,2) &=& \frac{E_{1,2}}{(2 \pi)^3} \, (2\pi R^2_s)^{3/2}
\,  (c_0 s_0) \, \exp\left[-\frac{R_{s}^2}{2} ({\mathbf
k}_1 + {\mathbf k}_2)^2\right] \nn \\
&+& \frac{E_{1,2}n_0^*}{(2 \pi)^3 } \, (2\pi \rho^2_*)^{3/2} \,
(2c_0 s_0) \exp{\left[-i \frac{m \langle u\rangle \rho_*^2
({\mathbf k}_1+{\mathbf k}_2)^2}{2 m_* RT} \right]}
\nonumber\\
&\times & \exp{\left[ - \frac{({\mathbf k}_1-{\mathbf k}_2)^2}{8
m_*T} - \frac{\rho_*^2}{2} ({\mathbf k}_1+{\mathbf k}_2)^2 -
\frac{(R^2+R_{_s}^2)({\mathbf k}_1+{\mathbf k}_2)^2} {8 m_* (R^2 T
+ R_{_s}^2 T_*)} \right]} .
\label{Gs12-2Vf}
\end{eqnarray}

The single-particle distribution for this situation of two volumes
can be written as
\bea
N^{2V}_1({\bf k}_i) &=& G^{2V}_c(i,i) \nn \\
&=& \frac{E_{i,i}}{(2 \pi)^3} \Bigl[ n^*_0 \left(|c_0|^2 +
|s_0|^2\right) I^c_{i,i}(R_s, R, T, m_*) + |s_0|^2
I^c_{i,i}(R_s, \infty, \infty, m_*) \nn\\
&+& n_0 I^c_{i,i}(\infty, R, T, m) - n_0 I^c_{i,i}(R_s, R, T,
m)\Bigr] \nn \\
&=&  \frac{E_{i,i}}{(2\pi)^3} \Biggl\{  (2\pi R^2_s)^{3/2} |s_0|^2
+ n_0^* \; (2\pi\rho^2_*)^{3/2} \left( |c_0|^2+|s_0|^2 \right)
\exp{\left[-\frac{(R^2 +R_{_s}^2){\mathbf k}_i^2} {2m_* (R^2 T +
R_{_s}^2 T_*)}\right]} \nonumber\\
&+& n_0 (2\pi {\tilde{\cal R}}^2)^{3/2} \exp{\left(-\frac{{\mathbf
k}_i^2}{2 m \tilde{\cal T}}\right)}  - n_0 (2\pi
{\tilde{\rho}}^2)^{3/2} \exp{ \left[-\frac{(R^2 +R{_s}^2){\mathbf
k}_i^2} {2m (R^2 T +R{_s}^2 \tilde{\cal T})}\right]} \Biggr\} .
\label{spectrum2Vf} \eea

Note again that in the large momentum region only the first term
survives, which corresponds to a constant contribution given by
the modified vacuum, and the size of this contribution is less
than in case of Eq. (\ref{spectrum1Vf}), as the in-medium modified
quanta do not fill the entire volume in the present case: $R_s <
R$.

\end{document}